\documentclass[nofootinbib,showpacs,aps,pra,twocolumn,floatfix,superscriptaddress]{revtex4}
\usepackage{amssymb,amsfonts}
\usepackage{graphicx}
\usepackage{bm}
\usepackage{amsmath}
\newcommand{\be}{\begin{equation}}
\newcommand{\e}{\end{equation}}
\newcommand{\beml}{\begin{subequations}}
\newcommand{\eml}{\end{subequations}}
\newcommand{\beq}{\begin{eqnarray}}
\newcommand{\eq}{\end{eqnarray}}
\newcommand{\ba}{\begin{array}}
\newcommand{\ea}{\end{array}}
\newcommand{\lt}{\left}
\newcommand{\rt}{\right}
\newcommand{\n}{\nonumber}
\newcommand{\la}{\langle}
\newcommand{\ra}{\rangle}
\newcommand{\tr}{{\rm Tr}\,}

\newcommand{\re}{\,{\rm Re}\,}
\newcommand{\ep}{\boldsymbol{\varepsilon}}
\newcommand{\unite}{\mathbf{\hat{e}}}

\newcommand{\Du}{\textbf{D}^{\dagger}}
\newcommand{\Dd}{\textbf{D}}
\newcommand{\vk}{\textbf{k}}

\newcommand{\tD}{\overleftrightarrow{\boldsymbol{\Delta}}}
\newcommand{\bra}[1]{\left|#1\right\rangle}
\newcommand{\ket}[1]{\left\langle#1\right|}

\newcommand{\stst}[1]{{\langle#1\rangle}_\text{ss}}
\begin{document}
\date{\today}
\title{Elastic vs. inelastic coherent backscattering of
 laser light by cold atoms:\\ a master equation treatment}

\author{Vyacheslav Shatokhin}
\affiliation{Max-Planck-Institut f\"ur Physik komplexer Systeme,
N\"othnitzer Str. 38, 01187 Dresden, Germany}
\affiliation{B.~I.~Stepanov Institute of Physics NASB, 220072
Minsk, Belarus}
\author{Cord A.~M\"uller}
\affiliation{Physikalisches Institut, Universit\"at Bayreuth, 95440
Bayreuth, Germany}
\author{Andreas Buchleitner}
\affiliation{Max-Planck-Institut f\"ur Physik komplexer Systeme,
N\"othnitzer Str. 38, 01187 Dresden, Germany}
\begin{abstract}
We give a detailed derivation of the master equation description of the
coherent backscattering of laser light by cold atoms. In
particular, our formalism accounts for the nonperturbative
nonlinear response of the atoms when the
injected intensity saturates the atomic transition. Explicit
expressions are given for total and elastic
backscattering intensities in the different
polarization channels, for the simplest nontrivial multiple scattering
scenario of intense laser light multiply scattering from two randomly placed atoms.
\end{abstract}
\pacs{
42.50.Ct, %Quantum optics (quantum description of atom-light interactions)
42.25.Dd, 32.80-t, 42.25.Hz
}
 \maketitle
\section{Introduction}

Localization phenomena in disordered systems have become a subject
of intense research \cite{local,Sheng}, since they highlight the
fundamental role of interference effects for wave propagation. A
prominent example is the coherent backscattering (CBS) of light in
dilute, disordered media, a hallmark of weak localization. CBS
manifests itself in the enhanced backscattering of the injected
radiation, due to the constructive interference between
time-reversed pairs of multiple-scattering amplitudes along a
given sequence of scatterers, which prevails even after disorder
averaging. This remarkable effect was observed for the first time
with light scattering from suspensions of polysterene particles
\cite{clasCBS}, but recently has also been reported for laser
light scattering from clouds of cold atoms
\cite{labeyrie99,katalunga03,chaneliere03}. With such quantum
scatterers -- possessing an internal electronic structure which
can be probed by the scattering light field in a controlled way
(e.g., by the appropriate choice of
     the laser frequency and/or of the atomic species) -- additional
     decoherence
processes are brought into play, which fundamentally affect the radiation
transport across the scattering medium. Elastic
Raman processes on degenerate atomic
transitions driven by the injected field change its polarization,
alike spin-flips of electrons scattering from
lattice impurities \cite{mueller02}. Inelastic processes are induced by
intense driving of the atomic transition, leading to its saturation and a
nonlinear response of the atom \cite{cohen_tannoudji}, manifest in the emission
of photons with frequencies different from the one injected.

Both these decoherence mechanisms reduce the CBS intensity, occur
in general simultaneously, and are experimentally very well
controlled. Hence, experiments on the multiple scattering of
coherent radiation on atomic scatterers provide an ideal testing
ground for the detailed analysis of coherent quantum transport in
disordered media, and of its sensitivity towards various sources
of decoherence. Furthermore, if we consider CBS and weak
localization as a precurser of strong (i.e., Anderson)
localization, decoherence phenomena affecting CBS are likely to
become detrimental for the latter. Anderson localization of light,
however, is an important experimental target, for fundamental as
well as for technological reasons. Consequently, beyond its
fundamental interest, a detailed theoretical and experimental
understanding of disorder- and/or decoherence-induced transport
phenomena is highly desirable for possible applications, which
currently emerge, e.g., in the area of random lasers \cite{cao}.

While the impact of elastic spin flip processes on the CBS signal
is nowadays well-understood, with quantitative accord between
experiment and theory \cite{labeyrie03}, nonlinear processes due
to the saturation of atomic transitions still challenge our
theoretical understanding. On the one hand, perturbative
approaches are -- by definition -- badly suited for the regime of
strongly driving intensities. On the other hand, exact solutions
which take into account arbitrarily high multiple-scattering
orders are prohibitive, due to the exponentially increasing number
of the contributing scattering paths and of the coupled internal
states of the atomic scatterers. Different approaches are
presently persued in the attempt to achieve a better understanding
of CBS in this parameter regime. These range from diagrammatic
techniques \cite{wellens04}, over Langevin equations
\cite{gremaud05}, to a master equation treatment
\cite{shatokhin05}, for a small number of atoms. In the present
paper, we give detailed account of the latter approach.

We will focus on the scenario set by the first experimental study
of saturation-induced effects on the CBS signal, performed with
cold Sr atoms \cite{chaneliere03}. In these experiments, the
injected laser was near-resonant with the $^1\!S_0\rightarrow
^{1}\!\!\!P_1$ transition, which has a nondegenerate ground state
and thus leaves no room for spin-flip processes. Consequently,
only inelastic scattering could cause decoherence and thus reduce
the CBS signal. This was indeed experimentally observed already
for moderate values of the atomic saturation parameter
\begin{equation}
s=\Omega^2/2(\delta^2+\gamma^2).
\label{satpar}
\end{equation}
$\Omega$ is the driving-induced Rabi frequency, $\gamma$ half the
spontaneous decay rate of the excited atomic level, and $\delta
=\omega_L-\omega_0$ the detuning of the injected laser frequency
$\omega_L$ from the exact atomic transition frequency $\omega_0$,
see Fig.~\ref{fig:levels}.

While our formalism to be unfolded hereafter is not restricted to the
treatment of atomic transitions with nondegenerate ground states, this
specialization allows for a more transparent presentation, and, in
particular, for a clear identification of the various inelastic processes
which intervene.

The paper is organized as follows: The next section starts out
with a general Hamiltonian formulation of the dynamics of $N$
atoms under coherent external driving, and coupled to the
electromagnetic vacuum. A master equation for the time evolution
of the atomic degrees of freedom constitutes the central building
block of the theory. Explicit expressions for the
(back-)scattering intensities in arbitrary polarization channels
are derived, in terms of the steady state quantum mechanical
expectation values of atomic dipoles and of dipole-dipole
correlation functions. Expansion of these to second order in the
dipole-dipole interaction constant between pairs of atoms finally
allows us to present analytic expressions for the
polarization-filtered backscattering signal, assuming that double
scattering processes provide the dominant contribution.
Accordingly, we restrict our final evaluation to the case of light
scattering from two, randomly placed atoms. Section~\ref{sec:conf}
provides a recipe of how to perform the disorder average, before
Sec.~\ref{sec:results} presents quantitative results for the
different polarization channels. Section~\ref{sec:summary}
concludes the paper.

\section{Master equation approach to coherent backscattering}
\label{sec:problem}

\subsection{Full $N$-atom master equation}
\begin{figure}
\includegraphics[width=6cm]{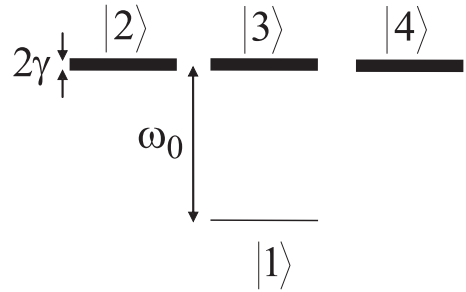}
\caption{
Level scheme of a $J_g=0\rightarrow J_e=1$ dipole transition,
with atomic transition frequency $\omega_0$, and natural linewidth $2\gamma$.
The sublevels $\bra 1$ and $\bra 3$ have magnetic quantum number
$m=0$. Sublevels $\bra 2$ and $\bra 4$ correspond to $m=-1$ and $m=1$,
respectively.
}
\label{fig:levels}
\end{figure}

We start with a general formulation of the Hamiltonian describing $N$
identical, motionless atoms with an isotropic dipole transition
coupled to the quantized photon reservoir
and driven by a quasiresonant (classical) laser field. The total
Hamiltonian of the system,
\be
H=H_{\rm A}+H_{\rm F}+H_{\rm AF}+H_{\rm AL},
\label{totalHamilt}
\e
contains the free atomic Hamiltonian $H_{\rm A}$, the free field Hamiltonian
$H_{\rm F}$, the atom-field coupling $H_{\rm AF}$, as well as the atom-laser
coupling $H_{\rm AL}$:
\beq
H_{\rm A}&=&\hbar\omega_0\sum_{j=\alpha}^N\Du_\alpha\cdot\Dd_\alpha,\\
H_{\rm F}&=&\hbar\sum_{{\bf k},s}\omega_k a^{\dagger}_
{{\bf k},s}a_{{\bf k},s},\\
H_{\rm AF}&=&\hbar\sum_{\alpha=1}^N\sum_{{\bf
    k},s}\left[\kappa^*_{\vk}({\bf r}_\alpha)
a^{\dagger}_{{\bf k},s}(\Dd_\alpha\cdot\ep_{{\bf
    k},s})\right.\n\\
&&\left.+\kappa_{\vk}({\bf r}_\alpha)
a_{{\bf k},s}(\Du_\alpha\cdot\ep^*_{\vk,s})\right],\\
H_{\rm AL}&=&-\frac{\hbar}{2}\sum_{\alpha=1}^N\left[\Omega_\alpha
e^{-i\omega_Lt}(\Du_\alpha\cdot\ep_L)\right.
\n\\
&&\left.+\Omega_\alpha^*e^{i\omega_Lt}(\Dd_\alpha\cdot\ep^*_L)\right]\,
. \eq Here, $\mathbf{D}^{(\dagger)}_\alpha$ is the lowering
(raising) operator for the isotropic dipole transition (see
Fig.~\ref{fig:levels}) at the resonance frequency $\omega_0$ of
atom $\alpha$, defined by \be
\Dd_\alpha=-\unite_{-1}\sigma^\alpha_{12}+\unite_0\sigma^\alpha_{13}-\unite_{+1}\sigma^\alpha_{14}\,
. \label{dipole} \e The $\sigma^\alpha_{kl}\equiv\bra k_\alpha\ket
l_\alpha$ mediate transitions between the electronic states of
atom $\alpha$,
%%%%%%%%%%%%%%%%%%%%%%%%%%%%%%%%%%%%%%%%
and \be \unite_{\pm 1}=\mp\frac{1}{\sqrt{2}}(\unite_x \pm i
\unite_y), \quad \unite_0=\unite_z \e are the unit vectors of the
spherical basis. In the free field Hamiltonian,
$a^{(\dagger)}_{{\bf k},s}$ annihilates (creates) a photon in the
reservoir mode with wavevector $\vk$ and transverse polarization
$\ep_{\vk,s}$, where $s$ is the polarization index.

The interaction Hamiltonians $H_{\rm AF}$ and $H_{\rm AL}$ are
written in rotating wave and dipole approximation. The coupling
constant between atom $\alpha$, located at point
$\textbf{r}_\alpha$, and the vacuum mode $(\vk,s)$ reads
\be
\kappa_{\bf k}({\bf r}_\alpha)=-id\lt(\frac{\omega_k}{2\hbar
\epsilon_0V}\rt)^{1/2}e^{i\vk\cdot{\bf r}_\alpha}\, , \label{kappa}
\e
where $d$ is a reduced matrix element, $\epsilon_0$ is the
permittivity
of the vacuum, and $V$ is the quantization volume. The coupling of
atom $\alpha$ to the laser field \be {\bf E}_L({\bf r})=\ep_L{\cal
E}e^{i(\vk_L\cdot{\bf r}-\omega_Lt)}+{\rm c.c.} \e is
characterized by a position-dependent Rabi frequency \be
\Omega_\alpha=\frac{2d{\cal E}}{\hbar}e^{i\vk_L\cdot{\bf
    r}_\alpha}\equiv\Omega e^{i\vk_L\cdot{\bf
r}_\alpha}\, .
\label{rabi}
\e

The figure of merit in our present study is the average value of
the stationary intensity $I({\bf r})$ with polarization $\ep $,
scattered in a direction close to backscattering $-\vk_L$: \be
I(\textbf{r})=\lim_{t\to\infty}\la[\ep\cdot{\bf
  E}^{(-)}({\bf r},t)][\ep^*\cdot{\bf E}^{(+)}({\bf r},t)]\ra \, .
\label{st_I} \e ${\bf E}^{(-/+)}({\bf r},t)$ is the
negative/positive frequency component of the source field
operators, given by the superposition of the retarded fields
radiated by all atomic dipoles, that is projected onto the
polarization vector $\ep$, upon detection, \be \ep^*\cdot{\bf
E}^{(+)}({\bf r},t) = \frac{\omega_0^2}{4\pi\varepsilon_0c^2r}
\sum_{\alpha=1}^N \ep^*\cdot \Dd_\alpha(t_\alpha)
e^{-i\vk\cdot{\bf r}_\alpha}\, , \label{Eplus} \e
with
$t_\alpha=t-|{\bf r}-{\bf r}_\alpha|/c$, and $\vk$ the wave vector
with the wave length of the injected laser radiation, pointing in
the observation direction (note that all the nontrivial spectral
information is contained in the time dependence of the atomic
dipole correlation function). This expression follows immediately
from generalizing familiar expressions for the far field radiated
by a single atom \cite{Carmichael} to the present case of an
atomic cloud, with the cloud's diameter much smaller than the
distance to the detector.

The
total scattered intensity is then obtained by inserting (\ref{Eplus})
and its conjugate into (\ref{st_I}), and reads, up to a prefactor,
\be I = \sum_{\alpha,\beta=1}^N
\stst{[\ep\cdot\Du_\alpha][\ep^*\cdot \Dd_\beta] }e^{i\vk\cdot{\bf
    r}_{\alpha\beta}}\, ,
\label{avint2}
\e
where `ss' stands for {\it steady state}, and ${\bf
r}_{\alpha\beta}\equiv {\bf r}_\alpha-{\bf r}_\beta$.

%%%%%%%%%%%%%%%%%%%%%%%%%%%%%%%%%%%%%%%%%%%%%%%%
\subsection{Coherent backscattering and polarization channels}

\begin{figure}
\includegraphics[width=8cm]{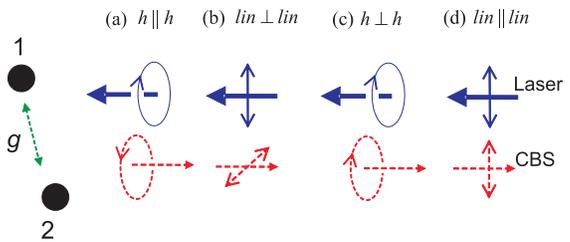}
\caption{(Color online) Elementary configuration for the
polarization-selective detection of the coherent backscattering
signal (CBS) (dashed arrows) from laser light (thick arrows)
scattering on two atoms 1 and 2 (black circles). The dipole-dipole
coupling strength between the atoms is given by the coupling
constant $g$. Backscattered photons are detected in four
polarization channels: (a) right circular in, left circular out
(with respect to a fixed observation direction; $h\parallel h$);
(b) linear in, orthogonal linear out ($lin \perp lin$); (c) right
circular in, right circular out ($h \perp h$); (d) linear in,
parallel linear out ($lin \parallel lin$). Note that in cases (c)
and (d) CBS appears on the background of single scattering from
independent atoms.} \label{fig:channels}
\end{figure}
%%%%%%%%%%%%%%%%%%%%%%%%%%%%%%%%%%%%%%%%%%%

The structure of Eq.~(\ref{avint2}), together with (\ref{dipole}), shows that
all three transitions between each atom's electronic levels (see
Fig.~\ref{fig:levels})
will in general
contribute to the scattered light intensity -- the relative weights
of their contributions depend on the observation direction
and on the detected polarization channel.

The unit polarization vector $\ep$ may be chosen
either in a circular or in a linear basis. For circular polarization, it is
convenient to use
the so-called helicity basis, in
which the quantization axis is directed along the probe direction
$\vk_L$.
In the case of linear polarization, the quantization axis is
conveniently chosen along the incident
polarization vector $\unite_0$ (perpendicular to $\vk_L$). For both
choices, $\ep$ has, in general,
three nonzero projections on the
unit vectors of the spherical basis, at finite angles between ${\bf k}$ and
$-\vk_L$.

From now on, we will use an approximation which is
legitimate at small scattering
angles ($\vk$ is very
close to $-\vk_L$), a common situation in CBS
experiments. One is interested in signals at very small angles
%CM:
$\theta \simeq 1/k\ell$ around the backscattering direction (the mean
free path $\ell$ is the average distance between consecutive
scatterers, and $k\ell\gg1$ in dilute atomic gases). Since the
geometric change of $\ep$ varies only with the cosine
of the scattering angle $\theta \ll 1$, we can  take $\ep$ constant,
equal to its value at exact backscattering.

With these conventions, Eq.~(\ref{avint2}) can be specialized for
the four polarization channels traditionally selected in CBS
experiments (see Fig.~\ref{fig:channels}).

\subsubsection{Helicity preserving channel $(h\parallel h)$}

In the helicity preserving channel, the
incident radiation is circularly
polarized, and drives either the $\bra 1\rightarrow\bra 2$ or the $\bra
1\rightarrow \bra 4$
transition. The backscattered light is then observed in the orthogonal
polarization channel (with the same helicity, since the propagation
direction is reversed) and must be radiated by the $\bra
1\rightarrow \bra 4$ or the $\bra 1\rightarrow\bra 2$ transition,
respectively. Both
combinations are completely equivalent, and for the case
$\ep_L=\unite_{+1}$,
$\ep=\unite_{-1}$, the total backscattered intensity reads
\be
I = \sum_{\alpha}\la\sigma_{22}^{\alpha}\ra_{\rm
ss}+\sum_{\alpha\neq\beta}\la\sigma_{21}^{\alpha}\sigma_{12}^{\beta}\ra_{\rm
  ss}e^{i\vk\cdot{\bf
r}_{\alpha\beta}}.
\label{inthparh1}
\e

\subsubsection{$lin \perp lin$ channel}

In the $lin \perp lin$ channel, $\ep_L=\unite_0$. We further
assume, without loss of generality, that the laser field is
propagating along the $x$-axis, such that the detected photons are
polarized along the $y$-axis. In the spherical basis,
$\ep=\unite_y=i(\unite_{-1}+\unite_{+1})/\sqrt{2}$, leading to the
expression \beq
I&=&\sum_{\alpha}\bigl(\la\sigma_{22}^{\alpha}\ra_{\rm
  ss}+\la\sigma_{44}^{\alpha}\ra_{\rm
ss}+\la\sigma_{24}^{\alpha}\ra_{\rm ss}+\la\sigma_{42}^{\alpha}\ra_{\rm
  ss}\bigr)\n\\
&&+\sum_{\alpha\neq\beta}e^{i\vk\cdot{\bf
r}_{\alpha\beta}}\bigr(\la\sigma_{41}^{\alpha}\sigma_{14}^{\beta}\ra_{\rm
ss}+\la\sigma_{21}^{\alpha}\sigma_{12}^{\beta}\ra_{\rm ss}\bigr.\n\\
\bigl.&&+\la\sigma_{41}^{\alpha}\sigma_{12}^{\beta}\ra_{\rm
ss}+\la\sigma_{21}^{\alpha}\sigma_{14}^{\beta}\ra_{\rm ss}\bigr).
\label{intlperl1}
\eq

\subsubsection{Flipped helicity channel $(h \perp h)$}

With incident
polarization $\ep_L=\unite_{+1}$ as before,
the $h\perp h$ channel corresponds to $\ep=\unite_{+1}$, such that
\be
I=\sum_{\alpha}\la\sigma_{44}^{\alpha}\ra_{\rm
ss}+\sum_{\alpha\neq\beta}\la\sigma_{41}^{\alpha}\sigma_{14}^{\beta}\ra_{\rm
  ss}e^{i\vk\cdot{\bf
r}_{\alpha\beta}}\, .
\label{inthperph1}
\e

\subsubsection{$lin \parallel lin$ channel}

Finally, in the  $lin \parallel lin$ channel,
the incoming and outgoing photons are
linearly polarized
along the same axis, $\ep_L=\ep=\unite_0$, and we obtain
\be
I=\sum_{\alpha}\la\sigma_{33}^{\alpha}\ra_{\rm
ss}+\sum_{\alpha\neq\beta}\la\sigma_{31}^{\alpha}\sigma_{13}^{\beta}\ra_{\rm
  ss}e^{i\vk\cdot{\bf
r}_{\alpha\beta}}\, .
\label{intlparl1}
\e

\bigskip

The above expressions systematically decompose into two parts:
Single atom, steady state dipole expectation values express the
intensities radiated by individual atoms, while correlation
functions of distinct atomic dipoles, multiplied by phases which
depend on the relative position of the atoms, account for the
interference contribution.

We now show how to evaluate these different steady state
expectation values explicitly, before performing the ensemble
average over the atomic positions, in Sec.~\ref{sec:conf}.

%%%%%%%%%%%%%%%%%%%%%%%%%%%%%%%%%%%%%%%%%%%%%%%%%%%%%%%%%%%%%%%%%%%%%%%%%%

\subsection{Equations of motion for the atomic correlation functions}

\label{subsection:eqs}

The dynamics of the atomic dipole operators' expectation values as
well as of the dipole-dipole correlators which enter
(\ref{inthparh1}-\ref{intlparl1}) is governed by the master
equation \cite{lehmberg} \be \la\dot Q\ra
=\sum_{\alpha=1}^N\la{\cal L}_\alpha Q\ra+\sum_{\alpha\neq
\beta=1} ^N\la{\cal L}_{\alpha\beta}Q\ra, \label{meq} \e where the
Liouvillians ${\cal L}_\alpha$ and ${\cal L}_{\alpha\beta}$
generate the time evolution of an arbitrary atomic operator $Q$,
for independent and interacting atoms, respectively, and
$\la\ldots\ra=\tr(\ldots\rho(0))$, with $\rho(0)$ being the
initial density operator of the $N$-atoms-field system, indicates
the quantum mechanical expectation value.

In the co-rotating frame with respect to the driving field at frequency $\omega_L$,
and
after the
standard electric dipole, rotating wave, and Born-Markov
approximations, ${\cal L}_\alpha$ and ${\cal
  L}_{\alpha\beta}$ read \cite{lehmberg}:
\begin{widetext}
\beq
{\cal L}_\alpha Q & = & -i\delta[\Du_\alpha\cdot\Dd_\alpha,Q]
-\frac{i}{2}[\Omega_\alpha(\Du_\alpha\cdot\ep_L)+\Omega^*_\alpha
(\Dd_\alpha\cdot\ep_L^*),Q]
+\gamma\lt(\Du_\alpha\cdot[Q,\Dd_\alpha]+[\Du_\alpha,Q]\cdot\Dd_\alpha\rt),\label{Lsingle}
\\
{\cal L}_{\alpha\beta}Q&=&\Du_\alpha\cdot\overleftrightarrow{\bf
    T}(g,{\bf\hat
n})\cdot[Q,\Dd_\beta]+[\Du_\beta,Q]\cdot\overleftrightarrow{\bf
T}^*(g,{\bf\hat n})\cdot\Dd_\alpha\, ,
\label{Liouvillians}
\eq
\end{widetext}
where the radiative
dipole-dipole
interaction due to
exchange of photons between the atoms is described by the tensor
$
\overleftrightarrow{\bf T}(g,{\bf\hat n})=\gamma g
\overleftrightarrow{\boldsymbol{\Delta}}$.
This interaction has a certain strength depending on the distance
between the atoms, via
\be
g =\frac{3i}{2k_0r_{\alpha\beta}}e^{ik_0 r_{\alpha\beta}},
\label{g}
\e
with $k_0=\omega_0/c$, and on the life time of the excited atomic levels,
through $\gamma$.

The coupling constant $|g|\ll 1$ is small
in the far-field ($k_0r_{\alpha\beta}\gg 1$), where near-field interaction
terms of order $(k_0r_{\alpha\beta})^{-2}$ and $(k_0r_{\alpha\beta})^{-3}$ can
be neglected
(which, at higher atomic densities, could also be retained in our formalism).
The projector $\overleftrightarrow{\boldsymbol{\Delta}}=\overleftrightarrow
{\openone}-{\bf \hat{n}\hat{n}}$ on the transverse
plane defined by
the unit vector $\bf\hat{n}$ along the connecting line between
the atoms $\alpha$ and
$\beta$ is explicitly given through
\beml
\beq
\overleftrightarrow{\openone}&=&-\unite_{-1}\unite_{+1}+\unite_{0}\unite_{0}-\unite_{+1}\unite_{-1},\\
{\bf\hat{n}}&=&\frac{e^{i\phi}\sin\vartheta}{\sqrt{2}}\unite_{-1}+\cos\vartheta\unite_0-
\frac{e^{-i\phi}\sin\vartheta}{\sqrt{2}}\unite_{+1}\, . \eq \eml The angles $(\vartheta,\phi)$,
which fix the direction of the connecting vector between two atoms
(with respect to the backscattering direction), will have to be
averaged over further down \cite{note}.

It should be kept in mind here that the master equation treatment
implies a trace over the modes of the free field, and that the
Markov approximation implies some coarse graining on the time
axis. Thus, $\gamma$ and $g$ are the only remnants of the coupling
to the electromagnetic vacuum, giving rise to some effective
dynamics of the atomic operators, on time scales which are long
with respect to the time scales of single absorption and emission
events from and into the electromagnetic reservoir. Only by
expansion of the solutions of (\ref{Lsingle},\ref{Liouvillians})
in powers of $g$ will we be able to distinguish multiple
scattering contributions of increasing order, since, formally, all
elastic and inelastic processes are lumped together in
(\ref{Lsingle},\ref{Liouvillians}) by $\Omega$, $\gamma$, and $g$.
This renders the master equation treatment somewhat less
transparent or at least less intuitive as compared to the
scattering theoretical approach \cite{wellens04}, but bears the
qualitative improvement of yielding results which are valid for
arbitrary saturation parameter $s$.

Note that the Markovian master equation
(\ref{meq}) ignores retardation effects
due to a
finite photon propagation time between scatters. This approximation is
justified when
$\max(r_{\alpha\beta}) \propto\ell \ll c/\gamma$ \cite{lehmberg}. In typical
experiments with sharply defined,
resonant optical dipole transitions \cite{labeyrie03}, both the condition of
diluteness, $k\ell\gg 1$, and of
`instantaneous'
propagation are very well satisfied.

The equation (\ref{meq}) leads to a system of linear, coupled
differential equations with constant coefficients for the atomic
correlation functions. The algebra of the $N$-atoms operators is
spanned by  tensor products of individual operators
$\sigma^\alpha_{kl}=\bra{k}_\alpha\ket{l}_\alpha$, each acting on
the $N$-fold tensor product of the four dimensional Hilbert space
in which the internal states of a single atom are represented. For
the free evolution of a single atom ($N=1$), $Q_\alpha$ can be
chosen in a complete orthonormal set of 16 basis operators (see
also Sect.~\ref{gmatrix} below, for details).

For our treatment of CBS, we need to include at least two-atoms
operators in (\ref{meq}). For $N=2$, the number of equations in
(\ref{meq}) is $255=16^2-1$ (there is one constant of motion). In
matrix notation, the resulting equation of motion reads \be
\dot{\la {\bf Q}\ra}=({\bf A}~+~{\bf V}){\la{\bf Q}\ra}+{\bf j}\,
, \label{matrixEq} \e where the elements of the vector $\la{\bf
Q}\ra$ are given by the expectation values of the complete
orthonormal set of two-atom operators. The elements $A_{nm}$,
$V_{nm}$ and $j_n$ of the matrices ${\bf
  A}$, ${\bf V}$,
and of the vector ${\bf j}$,
are derived from the equation for the
element $\la Q_n\ra$ in (\ref{matrixEq}):
\beq
\la({\cal L}_\alpha~+~{\cal L}_\beta)Q_n\ra&=&\sum_{m=1}^{255}A_{nm}\la
Q_m\ra+j_n,
\label{defAj}\\
\la({\cal L}_{\alpha\beta}~+~{\cal
  L}_{\beta\alpha})Q_n\ra&=&\sum_{m=1}^{255}V_{nm}\la Q_m\ra.
\label{defV}
\eq
From the decomposition of (\ref{matrixEq}) into (\ref{defAj}) and (\ref{defV})
it is apparent that the matrix ${\bf A}$ generates the evolution of uncoupled
atoms,
 whereas ${\bf V}$ describes their interaction via the exchange of
photons.

\subsection{Green's matrix and matrix ${\bf A}$}
\label{gmatrix}

In order to solve (\ref{matrixEq}), we first need its represention
in a suitable operator basis. Thereafter, we will derive a
solution for {\it independent} atoms (that is, we will ignore the
matrix ${\bf V}$, which mediates the interatomic correlations) by
a Laplace transform. The thus established relation between the
Green's matrix for the non-interacting two-atom system and the
matrix ${\bf A}$ will then serve as a basis for a systematic,
perturbative treatment of the interacting case, at increasing
order in the coupling constant $g$.

Let us first consider the dynamics of a single four-level system.
An expectation value of a single-atom operator $Q_\alpha$ obeys
the equation of motion \be \la\dot{Q_\alpha}\ra=\la{\cal L}_\alpha
Q_\alpha\ra, \label{meqQ} \e where the superoperator ${\cal
L}_\alpha$ is given by eq.~(\ref{Lsingle}). The master equation
(\ref{meqQ}) describes resonance fluorescence of the laser-driven
$J_g=0\rightarrow J_e=1$ atomic dipole transition. $Q_\alpha$
belongs to the complete orthonormal set $S^\alpha$ of 16 operators
for the four-level system, \be Q_\alpha\in
S^\alpha=\Bigl\{\frac{\openone^\alpha}{2},\frac{\mu_1^\alpha}{2},\frac{\mu_2^\alpha}{2},\frac{\mu_3^\alpha%%@
}{2},
\underbrace{\sigma^\alpha_{kl}\;(k\neq l=1\ldots 4)}_{12 \,{\rm
    operators}}\Bigr\}\, ,
\e
where
\beml
\label{s_i}
\beq
\openone&=&\sigma_{11}+\sigma_{22}+\sigma_{33}+\sigma_{44}\, ,\\
\mu_1&=&\sigma_{22}-\sigma_{33}+\sigma_{44}-\sigma_{11}\, ,\\
\mu_2&=&\sigma_{22}-\sigma_{33}-\sigma_{44}+\sigma_{11}\, ,\\
\mu_3&=&\sigma_{22}+\sigma_{33}-\sigma_{44}-\sigma_{11}\, . \eq
\eml It is easy to check that for the elements of $S^\alpha$ the
orthonormality condition $\tr[Q_nQ_m^T]=\delta_{nm}$ holds. In
this representation, equation (\ref{meqQ}) turns into a linear
matrix equation for the vector $\langle{\bf Q}_\alpha(t)\rangle$,
whose 16 elements are the quantum mechanical expectation values of
the elements of $S^\alpha$. Since the atomic levels' dynamics are
uncoupled, except for the laser-driven transition, it can be
solved analytically. The dynamics of the driven transition is
equivalent to the one of a two-level system.

For two atoms, $Q^{\alpha\beta}\in S^\alpha\otimes S^\beta$, and the vector of
the two-atoms correlation
functions,
\be
\langle{\bf
  Q}^{\alpha\beta}(t)\rangle=[\la\openone^\alpha\otimes\openone^\beta\ra/4,\ldots,
\la\sigma^{\alpha}_{kl}\otimes\sigma_{kl}^{\beta}\ra]^T \, ,
\e
consists of 256 elements. The evolution operator of {\it uncoupled} atoms reads
\be
e^{{\cal L}_0t}=e^{{\cal L}_\alpha t}\otimes e^{{\cal L}_\beta t}.
\label{Ltot}
\e
A Laplace transform $\int_0^\infty dt e^{-zt} e^{{\cal L}_0t}$ of (\ref{Ltot})
gives the Green's function (or resolvent) $G_{\alpha\beta}(z)=(z-{\cal
  L}_0)^{-1}$
of the Liouvillian
${\cal L}_0={\cal L}_\alpha+{\cal L}_\beta$.
In the two-atom basis $S^\alpha\otimes S^\beta$,
$G_{\alpha\beta}(z)$ has a matrix
representation ${\bf G}_{\alpha\beta}(z)$. This matrix has the following block
structure:
\be
{\bf G}_{\alpha\beta}(z)=
\begin{pmatrix}
z^{-1} &\vdots& {\bf 0}^T\\
\hdotsfor{3}\\
z^{-1}{\bf j}& \vdots&{\bf G}_0(z)
\end{pmatrix},
\label{matrG}
\e
where vectors ${\bf 0}$ (zero vector) and ${\bf j}$ have $255$ elements, and
${\bf G}_0(z)$ is the
truncated ($255\times 255$) Green's matrix. As seen from
(\ref{matrG}), the first column of matrix ${\bf G}_{\alpha\beta}(z)$ has a pole
at $z=0$. This pole
appears because the first element of the vector $\langle{\bf
  Q}^{\alpha\beta}(t)\rangle$, $\langle
\openone^{\alpha}\otimes\openone^{\beta}\rangle\equiv \tr\rho_0$, is a
constant of
motion. All other elements of the vector $\langle {\bf
  Q}^{\alpha\beta}(t)\rangle$ are time-dependent. The
steady-state solution of the truncated vector, $\langle{\bf Q}(t)\rangle$,
which
is obtained from $\langle
{\bf Q}^{\alpha\beta}(t)\rangle$ after exclusion of its first element,
$\tr\rho_0$, is defined as
$\langle{\bf Q}\rangle_{\rm ss}=\lim_{z\to 0}z\langle\tilde{{\bf
    Q}}(z)\rangle$, where
$\langle\tilde{{\bf Q}}(z)\rangle$ is the Laplacian image of the vector
$\langle{\bf Q}(t)\rangle$.
This
limit is evaluated to give
 \be
\langle {\bf Q}\rangle_{\rm ss}={\bf G}_0{\bf j},
\label{ssQ}
\e
where ${\bf G}_0\equiv {\bf G}_0(0)$.
%It is easy to see, by c
Comparison of
(\ref{ssQ}) with the steady-state
 solution of Eq.~(\ref{defAj}), $\langle {\bf Q}\rangle_{\rm ss}=-
{\bf A}^{-1}{\bf j}$, now shows that ${\bf G}_0\equiv -{\bf A}^{-1}$.

%%%%%%%%%%%%%%%%%%%%%%%%%%%%%%%%%%%%%%%%%%%%%%%%%%%%%%%%%%%%%%%%%%%%%%%%%%%%%%%

\subsection{Perturbative restriction to low scattering orders}
\label{perturbative}

The theoretical description of coherent backscattering is
relatively simple in two opposite regimes: either in the diffusive
regime of fully developed multiple scattering in optically thick
media, where long paths or high scattering orders yield the
celebrated conical line-shape of the CBS signal, or, on the
contrary, in the regime of scattering by optically thin media (or
in the presence of suppression of interference), where only double
scattering needs to be considered \cite{jonckheere00,bidel02}.
Indeed, it is in the double-scattering regime where the first
experimental observation of a CBS reduction due to the non-linear
saturation of atomic dipole transitions was reported
\cite{chaneliere03}. We limit our present analytical and numerical
analysis to this specific case.

In our master equation framework, all information on multiple
scattering processes is contained in the correlation functions of
dipole-dipole interacting atoms. The double scattering
contribution to the scattered light intensity from a given pair of
atoms, resulting from the exchange of two photons between the
atoms, is obtained by perturbative expansion of the respective
correlation functions to second order in the dipole-dipole
coupling constant $|g|$. This contribution depends only on the
observables related to the two selected atoms, and not on those of
the rest of the atoms in the cloud, since corrections to the mean
intensity due to the latter would be of higher order in $|g|$
($|g|^3$, etc.). Therefore, to find the double scattering
contribution, we will solve the master equation (\ref{meq}) for
two fixed atoms $\alpha=1,2$. Subsequently, we have to add up all
double scattering contributions resulting from the atoms located
at random positions. In other words, we need to perform
appropriate disorder averages of the solution for two fixed atoms.

This setup defines the simplest possible model describing CBS.
Indeed, double scattering is the lowest order process which gives
rise to distinct scattering amplitudes that can interfere
constructively. Despite its simplicity, this model allows for a
qualitative assessment of the impact of nonlinear scattering
processes on the CBS signal, whereas propagation effects in the
bulk of the scattering medium are beyond its reach. It also needs
to be considered that nonlinear scattering processes are induced
by high laser intensities, at which atoms are rapidly accelerated
out of resonance. Within our model, we neglect this acceleration,
by focusing exclusively on the coupling of photons to the internal
atomic degrees of freedom. Such an approximation is justified,
since the mechanical action of light on atoms can be
experimentally compensated by shortening the CBS probe duration,
as realized, e.g., in \cite{chaneliere03}.

For our perturbative solution of Eq.~(\ref{matrixEq}), we
take advantage of the small
parameter $g$ in (\ref{Liouvillians}) (defined in Eq.~(\ref{g})), and
expand in a power series of ${\bf V}$. The $n$th order
\be
\la{\bf Q}\ra^{[n]}_{\rm ss}=({\bf G}_0{\bf V})^n{\bf G}_0{\bf j}
\label{nth_power}
\e
of the stationary solution of Eq.~(\ref{matrixEq}) gives
the two-atoms correlation functions resulting from $n$ exchanged photons, and
generally
includes also recurrent scattering (a photon visits the same atom several
times). We
recall that, in the regime of elastic scattering from dilute
samples of resonant scatterers, higher scattering orders can be
accounted for by
considering more scatterers, while recurrent scattering is irrelevant
\cite{tiggelen96}.
In contrast, as one proceeds to the strong scattering regime, with denser
clouds of resonant scatterers,
the effect of recurrent scattering manifests by a
gradual reduction of the
 enhancement factor as compared to its maximum value $2.0$
\cite{wiersma95}. We will see
in Sect.~\ref{Totalhperph} below that Eq.~(\ref{nth_power}) generally including
recurrent scattering contributions is fully compatible
with neglecting these in the linear regime.

Total intensities for the double scattering contribution are given by
the third term (proportional to $|g|^2$) of the above expansion of the
correlation functions which enter
Eqs.~(\ref{inthparh1}-\ref{intlparl1}), with the general structure
\be
\la{\bf Q}\ra^{[2]}_{\rm ss}={\bf G}_0{\bf V}{\bf G}_0{\bf V}{\bf G}_0{\bf j}.
\label{doubleterm}
\e
Note that the matrix ${\bf V}$ depends on
$\overleftrightarrow{\bf\Delta}$, though neither on $\ep_L$ nor on
$\ep$. Indeed, information
about the laser polarization is carried by the Liouvillian (\ref{Lsingle}),
which governs the evolution of
independent atoms. As for $\ep$, it does not appear in the equations of motion
for the atomic correlation
functions at all -- but it defines which elements of the vector $\la{\bf
  Q}\ra^{[2]}_{\rm ss}$ contribute
to the observed signal, as evident from Eq.~(\ref{avint2}).

%%%%%%%%%%%%%%%%%%%%%%%%%%%%%%%%%%%%%%%%%%%%%%%%%%%%%%%%%%%%%%%%%%%

\subsection{Elastic component of double scattering}
\label{ecods}

The total backscattered intensity (\ref{avint2}) has a spectral distribution
that contains an elastic and, beyond the
weak field limit, also an inelastic component. The detected intensity is
the correlation function of the source field amplitudes radiated
by the atomic dipoles.

Its elastic component stems from the classically radiating
dipoles, i.e., from the nonfluctuating factorized averages
$\la\sigma_{i\neq j}^\alpha\ra$ \cite{cohen_tannoudji}. Hence, the
elastic intensity is given by the product of the expectation
values of the atomic dipoles, \be I^{\rm el} =
\sum_{\alpha,\beta=1}^N \stst{\ep\cdot\Du_\alpha}\stst{\ep^*\cdot
\Dd_\beta} e^{i\vk\cdot{\bf r}_{\alpha\beta}} \, .
\label{avintelastic}
\e
In the helicity preserving channel
$h\parallel h$, this reads \be I^{\rm el} =
\sum_{\alpha}|\la\sigma_{21}^{\alpha}\ra_{\rm
ss}|^2+\sum_{\alpha\neq\beta}\la\sigma_{21}^{\alpha}\ra_{\rm
  ss}\la\sigma_{12}^{\beta}\ra_{\rm
ss}e^{i\vk\cdot{\bf r}_{\alpha\beta}}.
\label{inthparh1elastic}
\e
Analogous expessions for the elastic component can be
derived from
Eqs.~(\ref{intlperl1}-\ref{intlparl1}) in the other channels.

The double scattering contribution $I^{\rm el}_2$ to the elastic
intensity, which we are interested in, and which is proportional
to $|g|^2$, is obtained from the corresponding power series
expansion, Eqs.~(\ref{nth_power},\ref{doubleterm}), of the
individual factors entering the above expression, in the coupling
constant. In this expansion, there emerge symmetric and asymmetric
combinations, like $\la\sigma_{21}^{1}\ra_{\rm
ss}^{[1]}\la\sigma_{12}^{2}\ra_{\rm ss}^{[1]}$, and
$\la\sigma_{21}^{1}\ra_{\rm ss}^{[2]}\la\sigma_{12}^{2}\ra_{\rm
ss}^{[0]}$, respectively. To ease the physical interpretation of
these various terms, remember that, by virtue of
(\ref{nth_power}), the superscripts `$[0]$', `$[1]$', and `$[2]$'
signal the scattering of a photon from one single atom $\alpha$
(`$[0]$'), subsequently from atom $\alpha$ and then from atom
$\beta$ (`$[1]$'), and the rescattering of the same photon from
atom $\alpha$, {\em after} a first encounter with $\alpha$ and
subsequent scattering from $\beta$ (`$[2]$').

In the $h \parallel h$ and $lin \perp lin$ channel, only symmetric
combinations contribute to the elastic component of the CBS
intensity, since the lowest order expectation value of a single
atom's coherence in the analyzed transition (like
$\la\sigma_{12}^{2}\ra_{\rm ss}^{[0]}$ in the helicity preserving
channel) must vanish: at lowest order, the atom is not coupled to
the other atom, nor is the transition directly driven by the
injected laser. By the same argument, also the single scattering
intensities from non-interacting atoms (arising from the first sum
on the right hand side of Eq.~(\ref{inthparh1elastic})) are
projected out, and an interference signal from purely multiple
scattering sequences is measured in these channels (unless the
atoms have a degenerate ground state and can undergo transitions
between different Zeeman-sublevels \cite{mueller01}).

In the $h \perp h$ and
$lin \parallel lin$ channels, both, symmetric and asymmetric combinations of
products of the
dipole averages contribute.

Altogether, at second order in the coupling constant, the elastic
backscattering intensity from two fixed atoms is obtained from the
evaluation of the zeroth-, first-, and second-order stationary
solutions $\la{\bf Q}\ra^{[0]}_{\rm ss}$, $\la{\bf
Q}\ra^{[1]}_{\rm ss}$, and $\la{\bf Q}\ra^{[2]}_{\rm ss}$ of
Eq.~(\ref{matrixEq}). The backscattering intensity from a cloud of
{\em randomly} located atoms is finally derived through an
appropriate disorder average.
%%%%%%%%%%%%%%%%%%%%%%%%%%%%%%%%%%%%%%%%%%%%%%%%%%%%%%%%%%%%%%%%%%%%%

\section{Disorder averaging}

%%%%%%%%%%%%%%%%%%%%%%%%%%%%%%%%%%%%%%%%%%%%%%%%%%%%%%%%%%%%%%%%%%%%%

\label{sec:conf}

Coherent backscattering is such a surprising effect
because it survives the ensemble average over random positions of the
scatterers, which
destroys all other speckle-like interferences.
The precise procedure of disorder
averaging is important as soon as one is interested in
the exact shape and angular width of the CBS cone. Here, we
rather focus on the impact of the atomic saturation on
the maximum CBS intensity, in the exact backscattering direction.
Since saturation
effects are independent of the precise averaging prescription, we
choose a procedure as simple as possible: an
(i) isotropic average of the relative orientation
${\bf\hat n}$ of the atoms
over the unit sphere is followed by (ii) an average of the
inter-atomic
distance $r_{12}$ over an interval of
the order of the laser wavelength, around their typical distance $\ell$:
\be
\la\ldots\ra_{\rm
  conf}=\frac{k_L}{4\pi}\int_{\ell-2\pi/k_L}^{\ell+2\pi/k_L}dr_{12}
\int d\Omega_{\bf\hat n}\ldots.
\label{conf}
\e

After the evaluation of (\ref{conf}), the final expression for the
backscattering intensity has a general structure which decomposes
into `ladder' and `crossed' contributions, respectively. The
ladder terms collect the intensities scattered by individual
atomic dipoles (in a diagrammatic representation, they arise from
the summation of co-propagating amplitudes \cite{Sheng} along a
sequence of scatterers), in an incoherent sum, whereas the crossed
terms stem from the interference of amplitudes radiated by
distinct dipoles (counter-propagating amplitudes in a diagrammatic
picture), and are garnished by the associated phases. Through this
phase factor, the interference part depends on the angle $\theta$
between the wavevector $\vk$ of the final photon and the
backscattering direction $-\vk_L$. As an example, the ladder and
crossed terms in the $h \parallel h$ channel read, by virtue of
Eq.~(\ref{inthparh1}), \beq L^{\rm
tot}_2&=&\la\la\sigma_{22}^1\ra^{[2]}_{\rm
  ss}+\la\sigma_{22}^2\ra^{[2]}\ra_{\rm ss}\ra_{\rm
conf},\label{Lterm}\\
C^{\rm tot}_2(\theta)&=&2\re\la\la\sigma^1_{21}\sigma^2_{12}\ra^{[2]}_{\rm
  ss}e^{i\vk\cdot{\bf
r}_{12}}\ra_{\rm conf}\, ,\label{Cterm}
\eq
respectively.

The enhancement factor $\alpha$, which is the figure of merit for the
quantification
of CBS, is given by
\be
\alpha=1 + \frac{C^{\rm tot}_2(0)}{L^{\rm tot}_2}\, .
\label{efactor}
\e
An analogous expression for the {\em elastic} CBS component follows from
Eq.~(\ref{inthparh1elastic}), with ladder and crossed
terms  $L^{\rm el}_2$ and
$C^{\rm el}_2(\theta)$.

%%%%%%%%%%%%%%%%%%%%%%%%%%%%%%%%%%%%%%%%%%%%%%%%%%%%%%%%%%%%%%%%%%%%%%%%%%%%%%%

\section{Results}
\label{sec:results}
%%%%%%%%%%%%%%%%%%%%%%%%%%%%%%%%%%%%%%%%%%%%%%%%%%%%%%%%%%%%%%%%%%%%%%%%%%%%%%

\begin{figure}
\includegraphics[width=7cm]{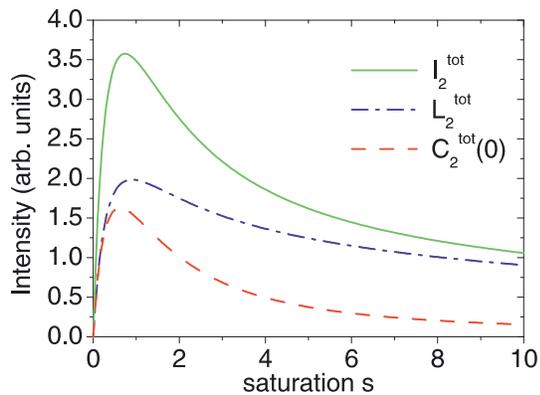}
\caption{(Color online) Saturation dependence of the total double
scattering contribution $I_2^{\rm tot}$ to the CBS
      signal in the $h \parallel h$ channel, decomposed in its ladder
      and interference parts
      $L_2^{\rm tot}$ and $C_2^{\rm tot}$, according to
Eqs.~(\ref{CTheta},\ref{LHparH}), at exact resonance $\delta=0$.
At finite saturation $s$, the interference term $C_2^{\rm tot}$
drops below the ladder contribution $L_2^{\rm tot}$, indicating a
loss of coherence. At large $s\ge 1$, the total double scattering
intensity must decrease with $s$, since the scattering cross
section of the emitting atom drops as $s^{-1}$.} \label{fig:hparh}
\end{figure}
%%%%%%%%%%%%%%%%%%%%%%%%%%%%%%%%%%%%%%%%%%%%%%%%%%%%%%%%%%%%

We now proceed to evaluate the general expressions derived above, for the four
typical polarization channels analyzed in the laboratory:
$h \parallel h$, $lin \perp lin$, $h \perp
h$, and $lin \parallel lin$. Analytical results for vanishing detuning $\delta$
will be complemented by some numerical results for $\delta\ne 0$.
%%%%%%%%%%%%%%%%%%%%%%%%%%%%%%%%%%%%%%%%%%%%%%%%%%%%%%%%%%%%%%%%%%%%%%%%%%%%%

\subsection{$h \parallel h$ channel}
\label{sec:r1}

\subsubsection{Total intensity, at zero detuning}
\label{totalintens,zerodet}

In Eqs.~(\ref{inthparh1}) and (\ref{inthparh1elastic}) we assumed
that the $\bra 1\rightarrow \bra 4$ transition is laser-driven.
Photons with preserved helicity originate from the  $\bra
2\rightarrow \bra 1$ transitions, and $\ep_L=\unite_{+1}$,
$\ep=\unite_{-1}$. The total double scattering intensity for two
fixed atoms reads, by virtue of Eqs.~(\ref{inthparh1}) and
(\ref{doubleterm}), at second order in $g$ and vanishing detuning
$\delta=0$:
\beq 2\re\{\la\sigma_{21}^1\sigma_{12}^2\ra_{\rm
ss}^{[2]}e^{i{\bf k}\cdot{\bf
r}_{12}}\}&=&|g|^2|\tD_{+1,+1}|^2\frac{R_1(s)}{(4+s)P(s)}\n\\
&&\times\cos\{({\bf k}+{\bf k}_L)\cdot{\bf r}_{12}\}\label{cohs},\\
\la\sigma_{22}^1\ra_{\rm ss}^{[2]}+\la\sigma_{22}^2\ra_{\rm
ss}^{[2]}&=&|g|^2|\tD_{+1,+1}|^2\frac{R_2(s)}{P(s)}\, . \label{pops}
\eq
$R_1(s)$, $R_2(s)$, and $P(s)$ are polynomial expressions in the
saturation parameter $s$,
\beml
\beq
R_1(s)&=&\frac{2}{9}\left(6912s+3168s^2\rt.\n\\
&&\lt.+264s^3+20s^4+s^5\rt),\\
R_2(s)&=&\frac{1}{3}\lt(1152s+528s^2+132s^3+7s^4\rt),\\
P(s)&=&(1+s)^2(12+s)(32+20s+s^2),
\eq
\label{RRP}
\eml
and
\be
\tD_{q,q'}\equiv \unite_q\cdot\tD\cdot\unite_{q^\prime}, \quad (q,q^\prime=\pm
1,0)\, .
\label{tD}
\e

The configuration average over (\ref{pops}) and (\ref{cohs}), defined in
Eq.~(\ref{conf}), leads to the final result
\beq
C^{\rm tot}_2(\theta)&\simeq&
\frac{|\tilde{g}|^2R_1(s)}{(4+s)P(s)}\Bigl(\frac{2}{15}-\frac{(k\ell\theta)^2}{35}\Bigr),\label{CTheta}\\
L^{\rm tot}_2&=&\frac{2|\tilde{g}|^2R_2(s)}{15P(s)},\label{LHparH}
\eq with $\tilde{g} = g|_{r_{\alpha\beta}=\ell}$ (see
Eq.~(\ref{g})). The scattering angle $\theta=2\arcsin\{|{\bf
k}+{\bf k}_L|/2k_L\}\ll 1$ with respect to the backscattering
direction was assumed to be sufficiently small herein. A power
series expansion of Eqs.~(\ref{CTheta},\ref{LHparH}) to second
order in $s$ reproduces the diagramatically obtained result of
\cite{wellens04}, \be C^{\rm tot}_2\propto s-\frac{5}{2}s^2, \quad
L^{\rm tot}_2\propto s-\frac{9}{4}s^2, \e for the total double
scattering intensity, at $\theta =0$.

The behavior of $L^{\rm tot}_2$, $C^{\rm tot}_2(0)$ shows that the
double scattering intensity $I^{\rm tot}_2= L^{\rm tot}_2+C^{\rm
tot}_2(0)$ behaves markedly different from that of an isolated
atom. While the radiated intensity from an isolated atom, \be
I^{[0]}\propto \frac{s}{1+s}, \label{I0} \e grows monotonically
with $s$ until it finally saturates \cite{cohen_tannoudji}, the
double scattering intensity exhibits a maximum at $s\simeq 0.7$
(see Fig.~\ref{fig:hparh}), followed by gradual decrease $\propto
s^{-1}$ for large $s$. Also this is a simple consequence of the
saturation behaviour of an isolated atom described by (\ref{I0}):
At high injected laser intensities, the atom that emits the final
photon has a total scattering cross section that asymptotically
decays like $I^{[0]}/I_L \propto s^{-1}$ (see
Eqs.~(\ref{satpar},\ref{rabi})), and is consequently less likely
to scatter photons coming from the other atom.

%%%%%%%%%%%%%%%%%%%%%%%%%%%%%%%%%%%%%
\begin{figure}
\includegraphics[width=7cm]{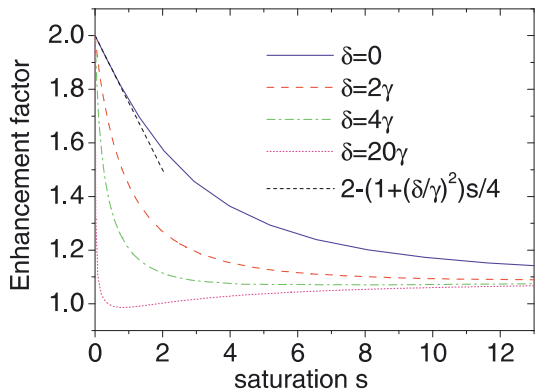}
\caption{
Numerical enhancement factor $\alpha$ in the helicity preserving channel $h
\parallel h$, versus
saturation $s$ of the atomic transition,
for different detunings $\delta$ from resonant driving. Larger detuning
leads to a faster loss of the CBS contrast. At small $s$, one
recovers the perturbative prediction \protect\cite{wellens04} linear in $s$
(straight dashed line). At large $s$, $\alpha$ saturates at a
$\delta$-dependent value $\alpha_{\delta_{\infty}}>1$, due to the constructive
self-interference of inelastically scattered photons. Remarkably, $\alpha$
also passes through a minimum at $s\simeq 0.5$, for very large detuning
$\delta =20\gamma$. This indicates {\em destructive} interference, and the
physical cause of this observation remains to be identified. }
\label{fig:enh}
\end{figure}
%%%%%%%%%%%%%%%%%%%%%%%%%%%%%%%%%%%%%%%%%%%%%%%%%%%%%%%%%%%%

The enhancement factor $\alpha(s)$, Eq.~(\ref{efactor}), deduced from
Eqs.~(\ref{CTheta},\ref{LHparH})
reads
\be
\alpha(s)=1+\frac{R_1(s)}{(4+s)R_2(s)}\, ,
\label{enh_res}
\e
and $\alpha(0)=2.0$ in the weak field limit, as
expected. The dependence of $\alpha$ on the saturation parameter is
shown in Fig.~\ref{fig:enh}.
As above for the individual cross and ladder terms, we again
obtain perfect agreement with the linear decay predicted by the
scattering theoretical result
$\alpha\simeq 2-s/4$ \cite{wellens04}, in the limit of small $s$.
When $s$ increases further, $\alpha$ monotonically drops to an
asymptotic value $\lim_{s\to\infty}\alpha(s)=\alpha_{\infty}=23/21$ which is strictly
larger than unity, implying a nonvanishing residual CBS contrast in
the limit of large injected intensities. As we shall see further down in
Sec.~\ref{sec:elastic_h_par_h},
this residual constructive interference effect is exclusively due to
the (self-)interference of {\em inelastically} scattered photons.

\subsubsection{Finite detuning}

It is in general no more possible to obtain explicit expressions
for the Green's matrix ${\bf G_0}$ (tantamount of inverting ${\bf
A}$), in the case of nonvanishing detuning $\delta\ne 0$. However,
this can always be done numerically, and Fig.~\ref{fig:enh}
compares the enhancement factor at resonance to the one for three
different nonvanishing values of $\delta$. In qualitative
agreement with the experiment \cite{chaneliere03}, $\alpha$ decays
faster for larger detuning, as $s$ is increased from zero. But not
only does the enhancement factor exhibit a steeper (initial)
decrease with $\delta$: it also reveals {\em destructive}
interference ($\alpha<1$) for large detuning $\delta=20\gamma$, at
$s\simeq 0.5$. This corresponds to a large Rabi frequency
$\Omega\simeq 20\gamma$.

\begin{figure}
\includegraphics[width=7cm]{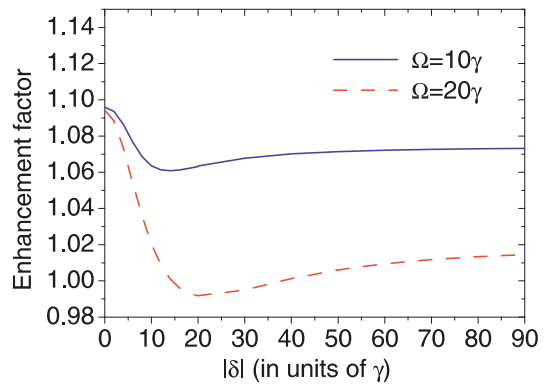}
\caption{(Color online) Numerical enhancement factor $\alpha$
(numerical solution) in the helicity preserving channel, versus
the laser detuning $\delta$, for two values of the driving Rabi
frequency $\Omega$. The decrease of $\alpha$ below unity for
$\Omega =20\gamma$, in the vicinity of $|\delta |=20\gamma$,
correlates with the minimum displayed by $\alpha(s)$   in the
corresponding plot in Fig.~\protect\ref{fig:enh}, at $s\simeq
0.5$. } \label{fig:detuning}
\end{figure}

To gain some insight
on whether such destructive interference is generic
for large
Rabi frequencies and large detunings, we monitor enhancement
factor vs. detuning, for two fixed, large values of the Rabi
frequency, as displayed in Fig.~\ref{fig:detuning}. For a given
value of $\Omega$, the enhancement factor decreases as a function
of $|\delta|$, from its maximum at $\delta=0$ to its
$\Omega$-dependent minimum value at $|\delta|\simeq \Omega$. For
very large detunings, $\alpha_\delta$ saturates at a level
$1<\lim_{|\delta|\to\infty}=\alpha_{\delta_\infty}<\alpha_\infty$,
indicating (i) constructive (self-)interference of far-detuned
photons, and (ii) similar behavior of the ladder and
crossed terms, asymptotically in $|\delta|$. The asymptotic value
$\alpha_{\delta_\infty}$ is the lower the larger $\Omega$.
Furthermore, as a direct counterpart of the destructive interference
observed in Fig.~\ref{fig:enh}, $\alpha$ drops below unity for
$\Omega=20\gamma$, in a finite range of $|\delta|$.

Note that a similar effect was predicted in \cite{Kupriyanov04},
for linear double scattering from atoms with Zeeman-shifted
hyperfine ground levels. In our case, the onset of destructive
interference at $\delta \simeq 15\gamma$ and $s_0 \simeq
\Omega^2/2\gamma^2= 200$ occurs approximately at saturation
$s\simeq 0.9$. A physical interpretation of this
interference-induced {\em anti-enhancement} of CBS, which we
tentatively attribute to an AC-Stark shift of the laser-driven
atomic sublevels, will require a closer inspection of the total
stationary intensity, and is refered to a separate contribution.

%%%%%%%%%%%%%%%%%%%%%%%%%%%%%%%%%
%%%%%%%%%%%%%%%%%%%%%%%%%%%%%%%%%%%%%%%%%%%%%%%%%%%%%%%%%%%%

\subsubsection{Elastic component at finite detuning}
\label{sec:elastic_h_par_h}

To see that the residual contrast observed in Fig.~\ref{fig:enh} for large
saturation parameters stems from inelastically scattered photons, we
now derive expressions for the elastic ladder and crossed
contributions to the double scattering CBS signal. To do so, we
extract the elastic contribution $I^{\rm
  el}_2$ to the total double scattering intensity from an expansion of
Eq.~(\ref{inthparh1elastic}) to second
  order in $|g|$, as prescribed by (\ref{nth_power},\ref{doubleterm}):
\be I^{\rm el}_2=|\la\sigma^1_{21}\ra^{[1]}_{\rm
  ss}|^2+|\la\sigma^2_{21}\ra^{[1]}_{\rm ss}|^2
+2\re\left (\la\sigma^1_{21}\ra^{[1]}_{\rm
  ss}\la\sigma^2_{12}\ra^{[1]}_{\rm ss}
e^{i{\bf k}\cdot{\bf
r}_{12}}\right )\, .
\label{el_corr_f}
\e

With the evolution equations \beml \beq
\dot{\la\sigma^\alpha_{12}\ra} & =
&(-\gamma+i\delta)\la\sigma^\alpha_{12}\ra-\frac{i}{2}\Omega_\alpha
\la\sigma^\alpha_{42}\ra\n\\
& & +\sum_{i,j}T_{ij}\la Q^\alpha_iQ^\beta_j\ra,
\label{Eq_sigma_12}\\
\dot{\la\sigma^\alpha_{42}\ra} & =
&-2\gamma\la\sigma^\alpha_{42}\ra-\frac{i}{2}
\Omega^*_\alpha\la\sigma^\alpha_{12}\ra\n\\
& & +\sum_{i,j}T_{ij}\la Q^\alpha_iQ^\beta_j\ra\, ,
\label{Eq_sigma_42} \eq \eml which can be derived from
Eqs.~(\ref{meq}-\ref{Liouvillians}), an analytic expression for
the steady state mean value $\la\sigma^\alpha_{12}\ra^{[1]}_{\rm
ss}$ and thus for $I^{\rm el}_2$ is obtained by the following
argument (note that this remains valid also for finite
detuning $\delta$): As long as we content ourselves with a lowest
order treatment of multiple scattering effects, only factorized
zeroth-order correlation functions $\la Q^\alpha_i\ra^{[0]}_{\rm
ss}\la Q^\beta_j\ra^{[0]}_{\rm ss}$ for independent atoms
$\alpha\ne\beta$ contribute to the sums on the rhs of
Eqs.~(\ref{Eq_sigma_12},\ref{Eq_sigma_42}), since the coefficients
$T_{ij}$ are of order $|g|$, expressing the dipole-dipole
interaction between distinct atoms. Such products vanish except if
their factors involve only the driven levels $\bra 1_\alpha$ or
$\bra 4_\beta$. Consequently, the summations in
Eqs.~(\ref{Eq_sigma_12}) and (\ref{Eq_sigma_42}), which extend
over different subsets of the two-atom correlation functions, can
be condensed according to \beq \sum_{i,j}T_{ij}\la
Q^\alpha_i\ra^{[0]}_{\rm ss}\la Q^\beta_j\ra^{[0]}_{\rm
ss}&\rightarrow&
\gamma g^*\tD_{+1,+1}\n\\
&&\times\la\sigma^\beta_{14}\ra^{[0]}_{\rm
  ss}\la\sigma^\alpha_{11}\ra^{[0]}_{\rm ss},\\
\sum_{i,j}T_{ij}\la Q^\alpha_i\ra^{[0]}_{\rm ss}\la
Q^\beta_j\ra^{[0]}_{\rm ss}&\rightarrow&
\gamma g^*\tD_{+1,+1}\n\\
&&\times\la\sigma^\beta_{14}\ra^{[0]}_{\rm
  ss}\la\sigma^\alpha_{41}\ra^{[0]}_{\rm ss}\, ,
\eq
respectively.
Substitution thereof into (\ref{Eq_sigma_12},\ref{Eq_sigma_42}) (with
the lhs of (\ref{Eq_sigma_12},\ref{Eq_sigma_42}) equal to zero, and
$\langle\sigma_{14}^{\alpha}\rangle =\langle\sigma_{41}^{\alpha}\rangle ^*$),
together with the known
solutions of the optical Bloch equations for a single
two-level atom \cite{cohen_tannoudji},
\be
\la\sigma^\alpha_{14}\ra_{\rm
  ss}^{[0]}=\frac{i(\gamma+i\delta)}{\Omega_\alpha^*}\frac{s}{1+s},\quad
\la\sigma^\alpha_{11}\ra_{\rm ss}^{[0]}=\frac{2+s}{2(1+s)}\, , \e
leads to \be \la\sigma^\alpha_{12}\ra^{[1]}_{\rm ss}=
\frac{i\gamma
g^*\tD_{+1,+1}\Omega_\beta}{2(\gamma-i\delta)^2(1+s)^2}\, .
\label{sigma_12} \e If we now rewrite Eq.~(\ref{el_corr_f}) as \be
I^{\rm el}_2=|T_{\rm dir}+T_{\rm rev}|^2, \label{sum_amplit} \e
with $T_{\rm dir}\equiv \la\sigma^1_{12}\ra^{[1]}_{\rm
ss}e^{-i{\bf
    k}\cdot{\bf r}_1}$
and $T_{\rm rev}\equiv \la\sigma^2_{12}\ra^{[1]}_{\rm ss}e^{-i{\bf k}\cdot{\bf
    r}_2}$, the elastic component of the double scattering intensity appears,
with (\ref{rabi}), as the square modulus of a sum of the `direct'
and `reversed' scattering amplitudes \beq T_{\rm
dir}&=&\frac{i\gamma
  g^*\tD_{+1,+1}\Omega}{2(\gamma-i\delta)^2(1+s)^2}e^{i{\bf k}_L\cdot{\bf
r}_2-i{\bf k}\cdot{\bf r}_1},\\
T_{\rm rev}&=&\frac{i\gamma
  g^*\tD_{+1,+1}\Omega}{2(\gamma-i\delta)^2(1+s)^2}e^{i{\bf k}_L\cdot{\bf
r}_1-i{\bf k}\cdot{\bf r}_2}\, .
\eq
Direct and reverse amplitude
are symmetric under the interchange ${\bf k}\leftrightarrow -{\bf
  k}_L$, and thus
satisfy the condition of reciprocity \cite{jonckheere00}.
Consequently, the elastically scattered photons remain strictly
coherent, for any $s$, and show perfect CBS contrast with
$\alpha(s) =2$, as immediately spelled out by the explicit
expression for the elastic ladder and crossed terms, at arbitrary
detuning and Rabi frequency: \be L^{\rm el}_2=C^{\rm
  el}_2(0)=\frac{2|\tilde{g}|^2}{15}\frac{1}{1+(\delta/\gamma)^2}\frac{s}{(1+s)^4}\, .
\label{coh_comp} \e
While perfectly coherent even for large $s$,
the elastic contribution to the total double scattering intensity
decreases as $s^{-3}$, by virtue of Eq.~(\ref{coh_comp}), after
passing through a maximum at $s=1/3$. In contrast, the total
signal fades away like $s^{-1}$, according to
Eqs.~(\ref{CTheta},\ref{LHparH}). Consequently, in the large $s$
limit, the CBS signal is completely dominated by the inelastic
scattering component, and the residual CBS contrast
$\alpha_{\infty}$ observed in Sec.~\ref{totalintens,zerodet} above
is due to the selfinterference of inelastically scattered photons,
which are incoherent with respect to the injected laser radiation.
The visibility of this residual interference signal is limited
by the amount of which-way information communicated to the
environment, during the multiple scattering process
\cite{Englert96}.

Finally, let us note that
equation (\ref{coh_comp}) allows for a transparent interpretation, since it can
be factorized
into
\begin{itemize}
\item[$(i)$]
the elastic intensity
\be
I^{\rm el[0]}\propto \frac{s}{(1+s)^2}
\e
scattered by the first strongly driven atom,
\item[$(ii)$]
the  total scattering cross section
\be
\sigma^{\rm tot}\propto \frac{1}{(1+(\delta/\gamma)^2)(1+s)}
\e
of the second atom, and
\item[$(iii)$]
the relative weight $I^{\rm el[0]}/I^{\rm tot[0]}=\sigma^{\rm
 el}/\sigma^{\rm tot}$ equal to
\be
\frac{\gamma^2+\delta^2}{\gamma^2+\Omega^2/2+\delta^2}=
\frac{1}{1+s}
\e
of elastic processes   \cite{cohen_tannoudji}.
\end{itemize}

%%%%%%%%%%%%%%%%%%%%%%%%%%%%%%%%%%%%%%%%%%%%%%%%%%%%%%%%%%%%%%%%%%%%%%%%5

\subsection{$lin \perp lin$ channel}
\label{sect:linperplin}

Up to a constant factor $1/2$, the results for the $lin \perp lin$
channel turn out to be the same as for the $h\parallel h$ channel,
at exact backscattering. For a given value of $s$, the ladder and
crossed terms are two times smaller than in the $h\parallel h$
channel. Since, however, Eq.~(\ref{intlperl1}) for the intensity
in the $lin \perp lin$ channel is manifestly different from
Eq.~(\ref{inthparh1}), the  $h \parallel h$ result, a short
discussion of this observation is in order.

In both cases, the CBS intensity is observed in the polarization
channel orthogonal to the excitation channel. In the $h\parallel
h$ channel, the orthogonal channel is defined by one dipole
transition $|2\ra\rightarrow|1\ra$. In the $lin\perp lin$ channel,
the orthogonal channel is defined by two atomic transitions,
$|2\ra\rightarrow|1\ra$ and $|4\ra\rightarrow|1\ra$. Yet, by
introducing a superposition state $|e\ra\equiv |2\ra+|4\ra$, we
can rewrite expression (\ref{intlperl1}) in a way which is
formally equivalent to (\ref{inthparh1}). However, the geometric
weight of the resulting ladder and crossed intensities is given by
$(\tD_{0,-1}+\tD_{0,+1})(\tD_{+1,0}+\tD_{-1,0})/2$. Only two of
these four terms, $\tD_{0,-1}\tD_{-1,0}/2$ and
$\tD_{0,+1}\tD_{+1,0}/2$, survive the configuration average, what
leads to results that are two times smaller than in the
$h\parallel h$ channel.

\subsection{$h \perp h$ channel}

We shall now consider detected photons which have the same
polarization as the incident ones. Hence, as already briefly
discussed in Sec.~\ref{ecods}, single scattering as well as double
scattering events will contribute to the detected signal, with the
latter only a small correction to the former. Correspondigly, an
experimental detection of the double scattering contribution alone
is excluded. Nonetheless, it is instructive to consider this
scenario in the regime of a nonlinear atomic response to the
injected radiation because it allows to identify the role of
recurrent scattering where one photon rescatters from the same
atom, after visiting the other (this process is second order in
the coupling constant $|g|$).

\subsubsection{Total intensity}
\label{Totalhperph}

To obtain explicit expressions for the total scattered intensity, we proceed
stepwise and first expand
the single atom contribution to the intensity, in Eq.~(\ref{inthperph1}), to
second
order in $g$. We obtain
\beq
\sum_{\alpha=1}^2\la\sigma_{44}^{\alpha}\ra_{\rm
ss}^{[2]}&=&|g|^2\left [F_1(s)|\tD_{+1,-1}|^2 \right .\label{incinthph}\\
&&\left.+F_2(s)(|\tD_{+1,0}|^2+|\tD_{+1,+1}|^2)\right ]\n\\
&&+\;{\rm terms}\;\propto \re[g^2]\, , |g|^2\cos\{2{\bf k}_L\cdot{\bf
  r}_{12}\}\, , \n
\eq
where
\beq
F_1(s)&=&\frac{36s+3s^2-27s^3-19s^4-s^5}{12(1+s)^5(3+s)}\, , \label{F1}\\
F_2(s)&=&-\frac{4s^2(288+132s+23s^2+s^3)}{3(1+s)P(s)}\label{F2} \,
, \eq with $P(s)$ from Eq.~(\ref{RRP}). The terms in the last line
of (\ref{incinthph}) oscillate rapidly on the typical scale $\ell$
of the interatomic separation (recall Eq.~(\ref{g}), and
$k_0r_{\alpha\beta}\gg 1$; this is, also the dependence of $g$ on
$r_{12}$ is to be taken into account here), and average out under
the integral over $r_{12}$ in (\ref{conf}). Thus, they will be
dropped hereafter, whereas terms $\propto |g|^2=9/4k_0^2r_{12}^2$
vary smoothly with $r_{12}$, and will be kept.

An analogous expansion of the interference terms in (\ref{inthperph1}) yields
the expression
\beq
2\re\{\la\sigma_{41}^1\sigma_{14}^2\ra_{\rm ss}^{[2]}e^{i{\bf k}\cdot{\bf
    r}_{12}}\}
&=&|g|^2F_3(s)|\tD_{+1,-1}|^2\label{naChph}\\
&&\times\cos\{({\bf k}+{\bf k}_L)\cdot{\bf r}_{12}\}\n\\
&&+\;{\rm terms}\;\propto \re[g^2]\, ,\n\\
&&\;\;\;|g|^2\cos\{({\bf k}-{\bf k}_L)\cdot{\bf r}_{12}\}\, ,\n\\
&&\;\;\;|g|^2\cos\{(3{\bf k}_L-{\bf k})\cdot{\bf r}_{12}\}\, ,\n
\eq
where
\be
F_3(s)=\frac{324s+540s^2+450s^3+219s^4+85s^5+29s^6+s^7}{36(1 +
  s)^6(3+s)^2}\, .
\label{D1}
\e
The three last lines of Eq.~(\ref{naChph}) are irrelevant for our
subsequent treatment, for exactly the
same reason
as the corresponding terms in Eq.~(\ref{incinthph}).

We now have a closer look at the geometric factors in these
equations, which allow the identification of the underlying
elementary scattering processes. The geometric weight of the
contribution proportional to $F_2(s)$ in (\ref{incinthph}) is
given by $|\tD_{+1,0}|^2+|\tD_{+1,+1}|^2$. It describes the
coupling of a photon from the $\bra 4_\alpha\rightarrow \bra
1_\alpha$ transition to either the $\bra 3_\beta\rightarrow \bra
1_\beta$, or to the $\bra 2_\beta\rightarrow \bra 1_\beta$
transition, respectively, then back  to the $\bra
4_\alpha\rightarrow \bra 1_\alpha$ transition, and only then to a
detector. This is recurrent scattering. (Note that non-recurrent
transitions, e.g., from $\bra 4_\alpha\rightarrow \bra 1_\alpha$
to $\bra 3_\beta\rightarrow \bra 1_\beta$ or to  $\bra
2_\beta\rightarrow \bra 1_\beta$ cannot give rise to a detected
photon in the $h\perp h$ channel.)

More precisely, the single scattering contribution with the weight
$F_2(s)$ originates from the interference between single
scattering from independent atoms and (recurrent) triple
scattering, in which a photon is subsequently scattered by atom
$\alpha$, then by atom $\beta$, and by atom $\alpha$ again (recall
our discussion in Sec.~\ref{ecods}, and also see
Eqs.~(\ref{Lelhph},\ref{elhperph2}) in our subsequent discussion
of the elastic contribution to the $h\perp h$ channel). Since the
rate of single and recurrent scattering is equally limited by the
number of photons incident on the atom, $\propto s$, it follows
that $F_2(s)\propto s^2$ for $s\rightarrow 0$. This is consistent
with a basic postulate of multiple scattering theory
\cite{tiggelen96}, according to which recurrent scattering is
irrelevant in the linear regime of weak saturation. It is also
clear why there is no recurrent scattering in $h\parallel h$ and
$lin \perp lin$ channels: Indeed, single scattering is essential
for recurrent scattering to show up at order $|g|^2$, but is
filtered out in the orthogonal polarization channels.

Now consider those terms in Eqs.~(\ref{incinthph}) and
(\ref{naChph}) with angular part $|\tD_{+1,-1}|^2$. Along with the
processes in which atoms exchange photons, these equally much
represent the two recurrent scattering sequences $\bra
4_\alpha\rightarrow \bra 4_\beta\rightarrow \bra 4_\alpha$. The
weight of these contributions is not the same as for $\bra
4_\alpha\rightarrow \bra 3_\beta\rightarrow \bra 4_\alpha$ and
$\bra 4_\alpha\rightarrow \bra 2_\beta\rightarrow \bra 4_\alpha$
transitions, since the presence of the driving field in the $\bra
1\leftrightarrow\bra 4$ transition definitely destroys the
symmetry of the excited state sublevels.

As regards the sign of the various scattering contributions in
Eqs.~(\ref{incinthph},\ref{naChph}), note that the weight of the
interference part given by Eq.~(\ref{D1}) is positive for all $s$,
while $F_1(s)$ and $F_2(s)$, Eqs.~(\ref{F1},\ref{F2}), are
nonpositive, the function $F_2(s)$ being strictly negative for
$s\neq 0$. Hence, recurrent scattering is a small negative
correction to the overall positive single-atom contribution to the
total scattering signal.

The total crossed and ladder contributions to the double
scattering intensity are once again obtained after a final
configuration average of Eqs.~(\ref{incinthph}) and
(\ref{naChph}):
\beq L^{\rm
tot}_2&=&\frac{1}{15}|\tilde{g}|^2(7F_1(s)+3F_2(s))\, ,
\label{Lhperpendh} \\
C^{\rm
  tot}_2(0)&=&\frac{7}{15}|\tilde{g}|^2F_3(s)\label{Crhperpendh}\,
.
\eq
%%%%%%%%%%%%%%%%%%%%%%%%%%%%%%%%%
\begin{figure}
\includegraphics[width=7cm]{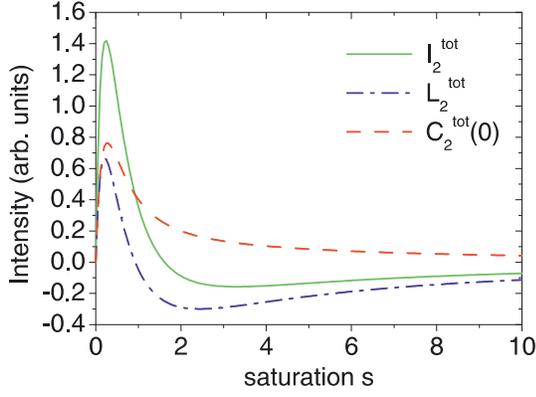}
\caption{(Color online) Total crossed (Eqs.~(\ref{Crhperpendh}))
and ladder (Eq.~(\ref{Lhperpendh})) terms, together with the
double scattering intensity $I_2^{\rm tot }=L_2^{\rm tot}+C_2^{\rm
tot}(0)$, in the $h \perp h$ channel, as functions of the
saturation $s$, at exact resonance $\delta =0$.}
\label{fig:hperph}
\end{figure}
%%%%%%%%%%%%%%%%%%%%%%%%%%%%%%%%%%%%%%%%%%%%%%%%%%%%%%%%%%%
The result is plotted in Fig.~\ref{fig:hperph}, where $L^{\rm
tot}_2$ turns negative in a finite interval of $s$. In the limit
of large $s$, the ladder term approaches zero from below, as
$1/s$, whereas the crossed terms decreases towards zero, with the
same rate $1/s$. Note that the negativity of the ladder term is
compensated for by the single scattering contribution $L_1$ which
cannot be separated from the double scattering contribution, in
the $h\perp h$ channel. Furthermore, the single scattering
contribution does not decrease with growing $s$ but rather
saturates, so that for very large saturation parameters we can
simply ignore the double scattering contribution. For all $s$, the
CBS enhancement factor reads, after inclusion of $L_1$:
\be
\alpha=1+\frac{C^{\rm tot}_2(0)}{L_1+L^{\rm tot}_2} \, .
\label{def2_enh}
\e

In a real medium, the relative weight of single and double
scattering depends on the optical thickness. However, our simple
model cannot correctly account for this effect; hence, we cannot
assess here whether the negativity of the double scattering ladder
term has observable consequences in laboratory experiments.

\subsubsection{Elastic component}

Let us finally extract the elastic component of the CBS intensiy
in the $h\perp h$ channel. According to (\ref{inthperph1}) and
(\ref{avintelastic}), with (\ref{conf}), the elastic ladder and
crossed terms are given by
\beq
L^{\rm
el}_2&=&2\lt\la|\la\sigma_{41}^1\ra^{[1]}_{\rm
  ss}|^2\rt. \label{Lelhph} \\
&&\lt.+\la\sigma^1_{41}\ra^{[0]}_{\rm ss}\la\sigma^1_{14}\ra^{[2]}_{\rm ss}+
\la\sigma^1_{41}\ra^{[2]}_{\rm ss}\la\sigma^1_{14}\ra^{[0]}_{\rm
  ss}\rt\ra_{\rm conf}\, , \n \\
C^{\rm el}_2(\theta)&=&2\re\lt\la e^{i\vk\cdot{\bf
    r}_{12}}\lt(\la\sigma^1_{41}\ra^{[1]}_{\rm
ss}\la\sigma^2_{14}\ra^{[1]}_{\rm ss}\rt.\rt. \label{elhperph2} \\
&&\lt.\lt.+\la\sigma^1_{41}\ra^{[0]}_{\rm ss}\la\sigma^2_{14}\ra^{[2]}_{\rm
ss}+\la\sigma^1_{41}\ra^{[2]}_{\rm ss}\la\sigma^2_{14}\ra^{[0]}_{\rm
  ss}\rt)\rt\ra_{\rm conf}\, . \n
\eq
Due to the factorization (\ref{avintelastic}) of the classically
  radiating dipoles, symmetric and asymmetric scattering contributions
  are directly born out: Products of first order (labeled by `$[1]$')
  contributions in $g$ represent double scattering of photons
  subsequently at
  atoms $1$ and $2$, in direct and reversed order, whereas products of
  zero and second order (`$[0]$' and `$[2]$', respectively) express
  indistinguishable single scattering and recurrent scattering
  amplitudes upon either one of the atoms.
Explicitly, the above expressions have the following geometric
weights,
\beq
L^{\rm el}_2&=&\lt\la|g|^2|\tD_{+1,-1}|^2\rt\ra_{\rm
  conf}\times\left [ \frac{s+s^3}{(1+s)^6} + F_4(s) \right ]  \n \\
&& + \lt\la|g|^2(|\tD_{+1,0}|^2+|\tD_{+1,+1}|^2)\rt\ra_{\rm
conf}\n\\
&&\times\frac{2F_2(s)}{1+s}\, , \\
C^{\rm el}_2(\theta)&=&\lt\la|g|^2|\tD_{+1,-1}|^2\cos\{({\bf k}+{\bf
  k}_L)\cdot{\bf
  r}_{12}\}\rt\ra_{\rm
conf} \n \\
&&\times \left [ \frac{s}{(1+s)^6} + F_5(s) \right ] \, \eq where
all terms which vanish under the configuration average have
  already been dropped (see also the discussion of
  Eqs.~(\ref{incinthph},\ref{naChph}) above), and
\beq
F_4(s)&=&\frac{-36s^2-39s^3-14s^4+s^5}{3(1+s)^6(3+s)}\, ,  \\
F_5(s)&=&\frac{-72s^2-51s^3-s^4+3s^5+s^6}{12(1+s)^6(3+s)}\, .
\eq

Upon evaluation of the configuration average, we obtain the final
result
\beq L^{\rm el}_2 &=& |\tilde{g}|^2\frac{7}{15} \left [
  \frac{s+s^3}{(1+s)^6} + F_4(s) + \frac{6}{7}\frac{F_2(s)}{1+s} \right ] \, ,
\label{l2el}
\\
C^{\rm el}_2(0) &=& |\tilde{g}|^2\frac{7}{15} \left [
  \frac{s}{(1+s)^6} + F_5(s) \right ]\, . \label{c2el}
\eq Expressions (\ref{l2el}) and (\ref{c2el}) imply that, in the
small-$s$ limit, the elastic ladder and crossed terms coincide, as
for orthogonal polarization channels (see Eq.~(\ref{coh_comp})).
However, this equipartition does {\em not} prevail here beyond the
linear regime, due to reciprocity violating processes
\cite{wellens04} which lead to a deviation of $L^{\rm el}_2$ from
$C^{\rm el}_2(0)$, already at quadratic order in $s$. The
violation of reciprocity was originally demonstrated in
\cite{wellens04} for scalar atoms, and it will be demonstrated
below in Sect.~\ref{sec:scalar} that the scalar results
immediately follow from our present results for parallel
excitation/detection polarization channels, when ignoring those
electronic sublevels that mediate recurrent scattering. However,
beyond those reciprocity-violating processes already implicit in
the scalar treatment, other processes specifically due to the
vector character of the injected radiation field lead to
additional deviations (expressed by the term $\propto F_2(s)$ in
(\ref{l2el})).

Furthermore, note that the elastic
ladder and crossed intensities in the $h\perp h$ channel asymptotically behave
 like $\propto
1/s^2$ and $\propto 1/s$, respectively. Also the total double scattering
intensity in this channel is characterized by an asymptotic
decrease $\propto 1/s$ (see Sec.~\ref{Totalhperph}). Consequently, unlike the
$h\parallel h$ channel,
where double scattering becomes purely inelastic for large $s$,
here both, elastic {\em and} inelastic photons, are present for large
$s$.

\subsection{$lin \parallel lin$ channel}

The results for the $lin \parallel lin$ channel are the same as
for the $h \perp h$ channel, modulo the following substitution of
the geometric weights: \be |\tD_{+1,-1}|^2\rightarrow
|\tD_{0,0}|^2, \quad |\tD_{+1,+1}|^2\rightarrow|\tD_{-1,0}|^2. \e
This entails slightly different final expressions for the ladder
and crossed terms. For the total intensities, we obtain \beq
L^{\rm tot}_2 &=& \frac{1}{15}|\tilde{g}|^2[8F_1(s)+2F_2(s)]
\label{Ltotpp}\, , \\
C^{\rm tot}_2(0) &=& \frac{8}{15}|\tilde{g}|^2F_3(s)\, ,
\label{Ctotpp} \eq and the elastic result reads \beq L^{\rm
el}_2&=&|\tilde{g}|^2\frac{8}{15}\Bigl[\frac{s+s^3}{(1+s)^6}+\frac{F_2(s)}{2(1+s)}+F_4(s)\Bigr]\,
, \label{Lsc1}\\
C^{\rm
  el}_2(0)&=&|\tilde{g}|^2\frac{8}{15}\Bigl[\frac{s}{(1+s)^6}+F_5(s)\Bigr]\label{Csc1}\, .
\eq
Eqs.~(\ref{Ltotpp},\ref{Ctotpp}) and (\ref{Lsc1},\ref{Csc1}) differ from
(\ref{Lhperpendh},\ref{Crhperpendh}) and
(\ref{l2el},\ref{c2el}) only through
numerical coefficients. Therefore, all our above conclusions for the
elastic and inelastic components of double scattering in the $h\perp
h$ channel also apply for the present $lin\parallel lin$ case.

\subsection{Scattering of scalar photons on a two-level atom}
\label{sec:scalar}
 To conclude, let us briefly consider the model scenario
of scalar photons scattering on a two-level atom -- a wide-spread
setting in typical quantum optical model calculations, which
neglects the important role of the polarization degree of freedom
in the presently discussed quantum transport problem. The scalar
case is easily deduced from the above results for the $h\perp h$
or $lin \parallel lin$ channels, by simply setting equal to zero
those contributions with the geometric weight
$|\tD_{+1,0}|^2+|\tD_{+1,+1}|^2$ -- since a two-level atom does
not offer the required atomic transitions.

The total ladder and crossed terms then read, by virtue of
Eqs.~(\ref{incinthph},\ref{naChph},\ref{Lhperpendh},\ref{Crhperpendh}),
\beq
L^{\rm tot}_2 &=& \frac{7}{15}|\tilde{g}|^2F_1(s)\, , \\
C^{\rm tot}_2(0) &=& \frac{7}{15}|\tilde{g}|^2F_3(s)\, , \eq with
the quadratic expansion \be L^{\rm tot}_2 \propto
s-\frac{21}{4}s^2\, ,\quad C^{\rm tot}_2(0)\propto s-5s^2\, , \e
in precise agreement with the result of \cite{wellens04}.

Correspondingly, the elastic contributions,
Eqs.~(\ref{l2el},\ref{c2el}), reduce to
\beq
L^{\rm
  el}_2&=&|\tilde{g}|^2\frac{7}{15}\Bigl[\frac{s+s^3}{(1+s)^6}+F_4(s)\Bigr]\, ,
\label{Lsc}\\
C^{\rm
  el}_2(0)&=&|\tilde{g}|^2\frac{7}{15}\Bigl[\frac{s}{(1+s)^6}+F_5(s)\Bigr]\, ,
\label{Csc} \eq which once again reproduces the second order
expression
\be
L^{\rm el}_2\propto s-10s^2,\quad C^{\rm
el}_2(0)\propto s-8s^2.
\e
of \cite{wellens04}.

Thus, the scalar model correctly predicts the maximum enhancement
factor $\alpha=2$, in the elastic scattering limit $s\to 0$.
Furthermore, it correctly describes those CBS contributions in the parallel
excitation/detection channels which originate
from the laser driven transitions, for arbitrary $s$. However, the scalar
model in general leads to incorrect results, since it ignores contributions
from those sublevels of the degenerate excited state
which are not driven by the laser, yet mediate recurrent
scattering, as we have seen in Sec.~\ref{Totalhperph}.

%%%%%%%%%%%%%%%%%%%%%%%%%%%%%%%%%%%%%%%%%%%%%%%%%%%%%%%%%%%%%%%%%%%%%%%%%%%%%%
\section{Summary}
\label{sec:summary}

In summary, we have given detailed account of the master equation
treatment of coherent backscattering of light from a disordered
sample of cold atoms with a nondegenerate electronic ground state.
This approach extracts all physical observables from the steady
state expectation values of atomic dipole operators. Furthermore,
our treatment incorporates, to the best of our knowledge for the
first time, the effect of interatomic dipole-dipole interactions
for {\em distant} atoms.

In particular, the formalism allows to treat arbitrary pump
intensities which possibly saturate the relevant atomic
transitions, thus leading to inelastic scattering events (when
more than one photon is incident on the scattering atom -- on the
spectral level, this entails the emergence of the famous Mollow
triplet, in a single atom's fluorescence). The price to pay is a
rapidly increasing dimension of the Hilbert space spanned by the
many atoms' degrees of freedom. This limited our present treatment
to two atomic scatterers, which is the minimum number of
constituents to observe the CBS effect. Nonetheless, this approach
allowed us to show that a small residual CBS signal survives even
in the limit of purely inelastic scattering, due to the
self-interference of inelastically scattered photons.

Furthermore, we have seen that recurrent scattering
leads to a reduction of the total double scattering signal, due to a
destructive interference between single and triple scattering events
upon the same atomic scatterer.

Another advantage of the master equation treatment presented here, so
far unexplored, is the immediate availability of the CBS {\em
  spectrum}
through a Fourier transform of suitable atomic dipole
correlation functions, as well as of the associated photocount
statistics. Whether CBS has an unambiguous signature in the spectrum
and/or in the photocurrent remains hitherto an open question, but is
getting in reach for state of the art experiments.

It is a pleasure to acknowledge entertaining and enlightening
discussions with Dominique Delande, Beno\^{\i}t Gr\'emaud,
Christian Miniatura, and Thomas Wellens.

%%%%%%%%%%%%%%%%%%%%%%%%%%%%%%%%%%%%%%%%%%%%%%%%%%%%%%%%%%%%%%%%%%%%%%%%%


\begin{thebibliography}{99}
\bibitem{note}
In the above, $\hat{n}\hat{n}$ and
  $\unite_{i}\unite_{j}$
have to be read as dyadic products.
\end{thebibliography}
\end{document}